\documentstyle[pre,aps,preprint,epsf]{revtex} \tightenlines
\begin{document} 

{\title{\bf Optical Limiting in Single-walled Carbon Nanotube
Suspensions}}

{\author{ S. R. Mishra, H. S. Rawat, S. C. Mehendale, K. C. Rustagi,
A. K. Sood{$^{\star}$},\\ Ranjini Bandyopadhyay{$^{\star}$},
A. Govindaraj $^{\dagger}$ and C. N. R. Rao{$^{\dagger}$}} \address {
Laser Programme, Centre for Advanced Technology, Rajendra Nagar,
Indore 452 013, India.}  \address{$^{\star}$ Department of Physics,
Indian Institute of Science, Bangalore 560 012, India.}
\address{$^{\dagger}$ Jawaharlal Nehru Centre for Advanced Scientific
Research, Jakkur P. O., Bangalore 560 064, India.}  \maketitle
\vspace{1cm} {\centerline{\it {Abstract}}}

Optical limiting behaviour of suspensions of single-walled carbon
nanotubes in water, ethanol and ethylene glycol is
reported. Experiments with 532 nm, 15 nsec duration laser pulses show
that optical limiting occurs mainly due to nonlinear scattering. The
observed host liquid dependence of optical limiting in different
suspensions suggests that the scattering originates from microbubbles
formed due to absorption-induced heating.

{\section{ Introduction}}

Carbon nanotubes provide a unique class of nanostructured
materials. Improved methods of synthesis, purification and
functionalization have triggered many experiments exploring basic
physics of mesoscopic objects as well as various possible applications
[1-4].  Like in C$_{60}$ and other fullerene derivatives, optical
limiting is an important application of nanotubes. Optical limiting in
liquid suspensions of multi-walled carbon nanotubes ( MWNTs ) has been
reported recently [5,6]. The observed limiting has been compared with
that of its well-studied cousin C$_{60}$ fullerene. Whereas optical
limiting in C$_{60}$ solution occurs due to the reverse saturable
absorption and subsequent nonlinear refraction and scattering [7],
optical limiting in MWNT suspension is attributed to the nonlinear
scattering arising from expanding microplasmas as in carbon black
suspension (CBS) [6]. Optical limiting investigations of carbon
nanotubes become even more important since they combine the relative
advantages of both CBS and fullerene solutions. As for CBS, the
nanotube suspensions consist of relatively large size (length $>$ 100
nm ) constituents. However, unlike CBS but like fullerenes, their
structure and electronic properties are well characterized and their
optical response is susceptible to further improvements by molecular
engineering.  We have carried out a detailed investigation of the
optical limiting of suspensions of well-characterised single-walled
carbon nanotubes ( SWNTs ) in ethanol, water and ethylene glycol by
carrying out optical limiting, z-scan and scattering measurements. The
results demonstrate that optical limiting in SWNT suspensions occurs
mainly due to absorption-induced scattering in the
suspension. Furthermore, the limiting strongly depends on the host
liquid. While this manuscript was in preparation, optical limiting in
a water suspension of SWNTs has been reported by Vivien et al
[8]. However, the present report includes studies of three host
liquids, an aspect relevant to optical limiting, and the results
suggest that the optical limiting is due to nonlinear scattering from
absorption-induced microbubbles.

\vspace{1cm} {\section { Experimental} }

SWNTs were produced by the dc arc discharge method using a composite
graphite rod containing Y$_{2}$O$_{3}$ (1 at.\%) and Ni (4.2 at.\%) as
anode and a graphite rod as cathode under a helium pressure of 660
torr with a current of 100 A and 30 V [9]. The web produced from the
arc-discharge contained SWNT bundles, amorphous carbon along with
metal encapsulated carbon particles as seen from the high resolution
electron microscope (HREM) image. It was heated in air at
300$^{\circ}$C for about 24 hours to remove the amorphous carbonaceous
materials. The heat-treated material was stirred with concentrated
nitric acid at 50$^{\circ}$C for about 12 h and washed with distilled
water to remove the dissolved metal particles. The SWNT material so
obtained was suspended in ethanol by using an ultrasonicator and
filtered through a micropore filter paper (0.3 $\mu$m) from Millipore
to remove polyhedral carbon nanoparticles present. The product was
then dried at 50$^{\circ}$C for about 12 h. The SWNT content of the
product was found to be 80\% by thermogravimetric
analysis. High-resolution electron microscopic examination (HREM)
showed that the SWNTs with an average diameter of 1.4 nm were present
as bundles of 10-50 nanotubes.  About 5-10 mg of the purified SWNT was
dispersed in 15 ml of water/ethanol by ultrasound sonication for 30
minutes. This dispersion was used for our study. All suspensions were
stable for several hours after ultrasonication. Optical transmission
spectra were found to remain unchanged during the experiment. Dynamic
light scattering measurements were performed on the suspensions to
characterise the average size of the scatterers. The light scattering
experiments were done using $\lambda$ = 647.1nm radiation from a
Kr$^{+}$ ion laser, a home-made spectrometer and a correlator (Malvern
7132CE 64 channel model). Fig. 1 shows the plot of the normalised
intensity autocorrelation function g$_{2}$(t) - 1= $<$I(0)
I(t)$>$/$<$I(0)$>$$^{2}$ - 1 versus time for different scattering
angles $\theta$, which have been fitted to g$_{2}$(t) - 1 = A exp
(-$\Gamma$ t ), where $\Gamma$ = 2Dq$^{2}$, q =$\frac{4\pi
n}{\lambda}$ sin$\frac{\theta}{2}$ is the wavevector transfer, n =
1.33 is the refractive index of the solvent and D is the diffusion
coefficient of the scatterer (nanotube). The inset of Fig. 1 shows a
plot of $\Gamma$ versus q$^{2}$, and a linear fit gives D =
1.49*10$^{-8}$ cm$^{2}$s$^{-1}$. Taking the average diameter of the
nanotube bundles to be 40nm as suggested by electron microscopy and
using the equations D = k$_{B}$T [6 - 0.5($\gamma_{\parallel}$+
$\gamma_{\perp}$)]/3$\pi \eta_{\circ}$L, $\gamma_{\parallel}$ = 1.27 -
7.4($\frac{1}{\delta}$ - 0.34)$^{2}$, $\gamma_{\perp}$ = 0.19 -
4.2($\frac{1}{\delta}$ - 0.39)$^{2}$ and $\delta$ = ln $\frac{2L}{d}$
[10], where L and d are the length and diameter of the cylindrical
rods and $\eta_{\circ}$ is the solvent viscosity, we obtain an average
value for L $\sim$ 25$\mu$m.  Experiments were performed using a
frequency doubled Nd:YAG laser giving 532 nm, 15 ns laser pulses with
pulse repetition every $\sim$3 seconds.  The laser emission was
focussed using a 50 cm lens such that 1/e$^{2}$ radius of the focussed
beam was $\sim$ 50 $\mu$m at the focus. For optical limiting
measurements the nanotubes sample was kept at the focus and the
transmitted emission was passed through an aperture with $\sim$ 95\%
transmission at low fluences. Thus the aperture would block off-axis
scattered emission, if any, from the sample at high fluences. The
input energy was measured by taking a fraction of the input beam on a
calibrated biplanar photodiode and output energy was measured by a
calibrated PIN-photodiode kept after the aperture. The input and
output fluences were estimated by measuring input and output energy
and beam size at the sample position.  \vspace{1cm} {\section{ Results
and Discussion}}
 
Fig. 2 shows the observed variation of output fluence with input
	fluence for a SWNTs suspension in water and a C$_{60}$
	-toluene solution . The transmission spectrum of SWNT sample
	is shown in the inset. Both absorption and scattering may
	contribute to the loss in transmitted intensity. The SWNTs and
	C$_{60}$ samples were kept in two identical cuvettes and
	concentrations were adjusted so that the low fluence
	transmission through the cuvette was the same ( $\sim$ 55 \% )
	for both the samples. It is evident from the figure that
	optical limiting in C$_{60}$ solution is stronger than that in
	SWNTs suspension in water. On the other hand, stronger
	limiting in MWNTs suspension in ethanol than that in C$_{60}$
	solution in toluene has been earlier reported by Chen et al
	[6].  To know the contribution of nonlinear refraction to the
	observed optical limiting in nanotubes suspension, we
	performed z-scan measurements [11]. In these measurements the
	aperture used in the limiting geometry was moved to the
	far-field region with $\sim$ 16\% transmission without the
	nanotubes sample in the beam path. Fig. 3 shows the z-scan
	results. The transmission shown in this figure is the ratio of
	the aperture transmission measured with the sample in the beam
	path to that measured without the sample. Thus at large z
	values the transmission corresponds to the low intensity
	transmission through the sample. At smaller z values ( i.e
	higher fluences ) transmission reduces and reaches its minimum
	near z=0 . The absence of any peak in the z-scan data in
	Fig. 3 indicates that nonlinear scattering is much stronger
	than nonlinear refraction. The observed pronounced valley
	shown in this figure can occur due to nonlinear scattering as
	well as nonlinear absorption. We believe that the dominant
	contribution to the observed transmission valley in Fig. 3 is
	from nonlinear scattering because the optical limiting was
	found to be very sensitive to the size of the aperture. We
	note that optical limiting in multi-walled carbon nanotubes
	suspensions has also been attributed to absorption induced
	scattering in the suspensions [6]. Z-scan measurement of SWNT
	suspension in water by Vivien et al [8] showed negative
	lensing corresponding to thermally induced nonlinear
	refraction, in apparent variance with our observations. The
	difference could be due to much higher intensities and much
	smaller aperture transmission used by them. On the other hand,
	Chen et al [6] found stronger optical limiting in MWNT
	suspension in ethanol compared to C$_{60}$ solution in
	toluene. Similarly, Vivien et al [8] also reported stronger
	optical limiting in SWNT suspension in water compared to
	C$_{60}$ solution in toluene. We note that optical limiting
	due to nonlinear refraction of thermal origin would be
	stronger for longer pulses used in our experiments. This
	mechanism is expected to be more important in C$_{60}$
	solution than in SWNT suspensions. To understand the nature of
	scattering, we measured the scattered light at different
	angles from the sample cell. Fig. 4 shows the results for an
	angle $\sim$ 0.8$^{\circ}$ from the beam axis. For these
	measurements, the energy of scattered light passing through a
	suitably positioned 3 mm diameter aperture was measured using
	a PIN photodiode in integrating mode. The y-axis in Fig. 4
	shows the ratio of the signal from this PIN photodiode to the
	signal from another photodiode monitoring the input pulse
	energy. For linear scattering this ratio is expected to be
	constant for all values of the input fluence. The data in
	Fig. 4 clearly shows evidence of nonlinear scattering. The
	ratio initially rises and then falls as input fluence is
	increased which appears to be due to a change in angular
	distribution of scattered light.  We have also measured
	optical limiting in nanotubes suspended in different liquids
	viz. ethylene glycol, water and ethanol. For this purpose,
	identical 10 mm path length cuvettes were used and low fluence
	transmission was kept at $\sim$ 42\%. During measurements with
	the three samples, all the cells containing SWNTs suspension
	were accurately positioned at the same place and other
	experimental set-up was unaltered. Fig. 5 shows the observed
	limiting data for different suspensions. It is evident that
	optical limiting in ethanol suspension was strongest among the
	three suspensions. The threshold fluence (defined as the
	fluence at which the transmission reduces to half of its low
	fluence value ) in different liquids was also different. As
	estimated from Fig. 5, the lowest value was $\sim$ 1.0
	J/cm$^{2}$ for ethanol suspension. These observations suggest
	that the extent of nonlinear scattering depends on the liquid
	host.  The nonlinear scattering from suspensions of absorbing
	particles in liquids can result from several mechanisms . The
	scattering centres can be expanding microplasmas generated by
	vaporization of the particles [12], bubbles formed by
	vaporization of the liquid [13] or transient refractive index
	inhomogeneities resulting from localised heating around
	absorbing particles [7]. Our observation of strong solvent
	dependence of optical limiting rules out plasma formation as
	being an important mechanism. The change in refractive index
	of a liquid depends on its thermal figure of merit F =
	$\frac{1}{c \rho} \frac{dn}{dT}$ (where $c$ is the specific
	heat, $\rho$ is the density and $\frac{dn}{dT}$ is the
	thermo-optic coefficient).  The values of F are 7.5, 8.0 and
	1.0 10$^{-4}$cm$^{3}$calK$^{-1}$ for ethylene glycol, ethanol
	and water , respectively. We note that although ethylene
	glycol has much larger F, optical limiting in glycol
	suspension was much weaker than that in water suspension. This
	implies that absorption induced refractive index
	inhomogeneities did not make significant contribution to the
	nonlinear scattering. We thus believe that the observed
	nonlinear scattering in our experiments is mainly from
	microbubbles formed in the suspension. The strongest optical
	limiting in ethanol suspension is qualitatively consistent
	with its lowest boiling point of $\sim$ 78 $^{\circ}$C among
	the three liquids. This conclusion also agrees with the recent
	pump-probe investigation of CBS by Durand et al [14] using 30
	ps laser pulses. These authors found that while for the first
	few nanosecond after the pump pulse, probe attenuation was
	solvent independent suggesting scattering from microplasma,
	for larger delays the probe transmission showed strong solvent
	dependence implying scattering from bubbles or thermo-optic
	effects.  In conclusion, we report optical limiting of visible
	ns laser pulses in suspensions of single-walled carbon
	nanotubes. The dominant mechanism for the observed limiting
	has been found to be absorption induced nonlinear
	scattering. Optical limiting has been found to be
	significantly different for different liquids indicating that
	scattering is probably due to bubble formation in the
	suspension, although other causes like sublimation of
	particles cannot be ruled out at present.
\vspace{1cm} {\section{ Acknowledgements}}

	AKS thanks Department of Science and Technology, New Delhi for
	financial assistance. Technical help of S. K. Tiwari and
	M. Laghate during experiments on optical limiting is
	gratefully acknowledged.

\vspace{1cm} {\section {References}} [1] S. Ijima, Nature 354, 56
(1991).
\newline
[2] T. W. Ebbesen, "Synthesis and Characterization of Carbon
Nanotubes" in Physics and Chemistry of Fullerenes, Ed., K. Prassides (
Kluwer Academic Publishers, Netherlands, 1994), T. W. Ebbesen, "Carbon
Nanotubes", Physics Today ( June 1996, p-26 ).
\newline
[3] C.N.R.Rao, A. Govindaraj, R.Sen and B. Satishkumar,
Mater. Res. Innovat 2, 128 (1998) and references therein.
\newline
[4] R. Saito, G. Dresselhaus and M. S. Dresselhaus, "Physical
Properties of Carbon Nanotubes" ( Imperial College Press, London, 1998
)
\newline
[5] X. Sun, R. Q. Yu, G. Q. Xu, T. S. A. Hor and W.Ji ,
           Appl. Phys. Lett. 73, 3632 (1998).
\newline
[6] P. Chen, X. Wu, X. Sun, J. Lin, W. Ji and K. L. Tan,
Phys. Rev. Lett. 82, 2548 (1999).
\newline
[7] S. R. Mishra, H. S. Rawat, M.P. Joshi and S. C. Mehendale,
    J. Phys. B 27, L157 (1994).
\newline
[8] L. Vivien, E. Anglaret, D. Riehl, F. Bacou, C. Journet, C. Goze,
     M. Andrieux, M. Brunet, F. Lafonta, P. Bernier, F. Hache,
     Chem. Phys. Lett. 307, 317 (1999).
\newline
[9] Journet, C., Maser, W. K., Bernier, P., Loiseau, A., Lamy de la
     Chapelle, M., Lefrant, S., Denierd, P., Lee, R. and Fischer,
     J. E., Nature, 1997, 388, 756.
\newline
[10] S. Broersma, J. Chem. Phys. 32, 1626 (1960).
\newline
[11] M. Sheik-Bahae, A. A. Said, T. H. Wei, D. J. Hagen and E. W. Van
    Stryland, IEEE J.Quant. Electr. 26, 760 (1990).
\newline
[12] K. Mansour, M. J. Soilcau and E. W. Van Stryland,
     J.Opt. Soc. Am. B 9, 1100 (1992).
\newline
 [13] T. F. Bogges, G. R. Allan, D. R. Labergerie, C. H. Venzke, A. L
 Smirl, L. W. Tutt, A. R. Kost, S. W. McCahahon, M. B. Klein,
 Opt. Engg. 32(5), 1063 (1993)
\newline
[14] O. Durand, V. Grolier-Mazza and R. Frey, Opt. Lett. 23, 1471
(1998).
\vspace{1cm} {\section {Figure Captions}}

Fig. 1.  Intensity autocorrelation functions measured at different
scattering angles in dynamic light scattering experiments with SWNT
suspension in water. The inset shows the linear fit to the plot of
$\Gamma$ versus q$^{2}$.\\

Fig. 2. Optical limiting in SWNTs - water suspension and
C$_{60}$-toluene solution. Circles show output with SWNTs-water
suspension and triangles show output with C$_{60}$ solution. The inset
shows transmission loss spectrum of SWNTs-water suspension.\\

Fig. 3. Variation of transmission with z in a close-aperture z-scan
measurements on SWNTs-water suspension. \\

Fig. 4. Measured variation of ratio of the scattered energy to the
input fluence with input fluence at an angle of $\sim$ 0.8$^{\circ}$
in the forward direction from the beam axis.\\

Fig. 5. Measured variation of output fluence with input fluence in
SWNTs suspensions in different solvents. Crosses, circles and
triangles show output from SWNTs suspension in ethanol, water and
ethylene glycol respectively.\\

\begin{figure}
\centerline{\epsfxsize = 8cm \epsfbox{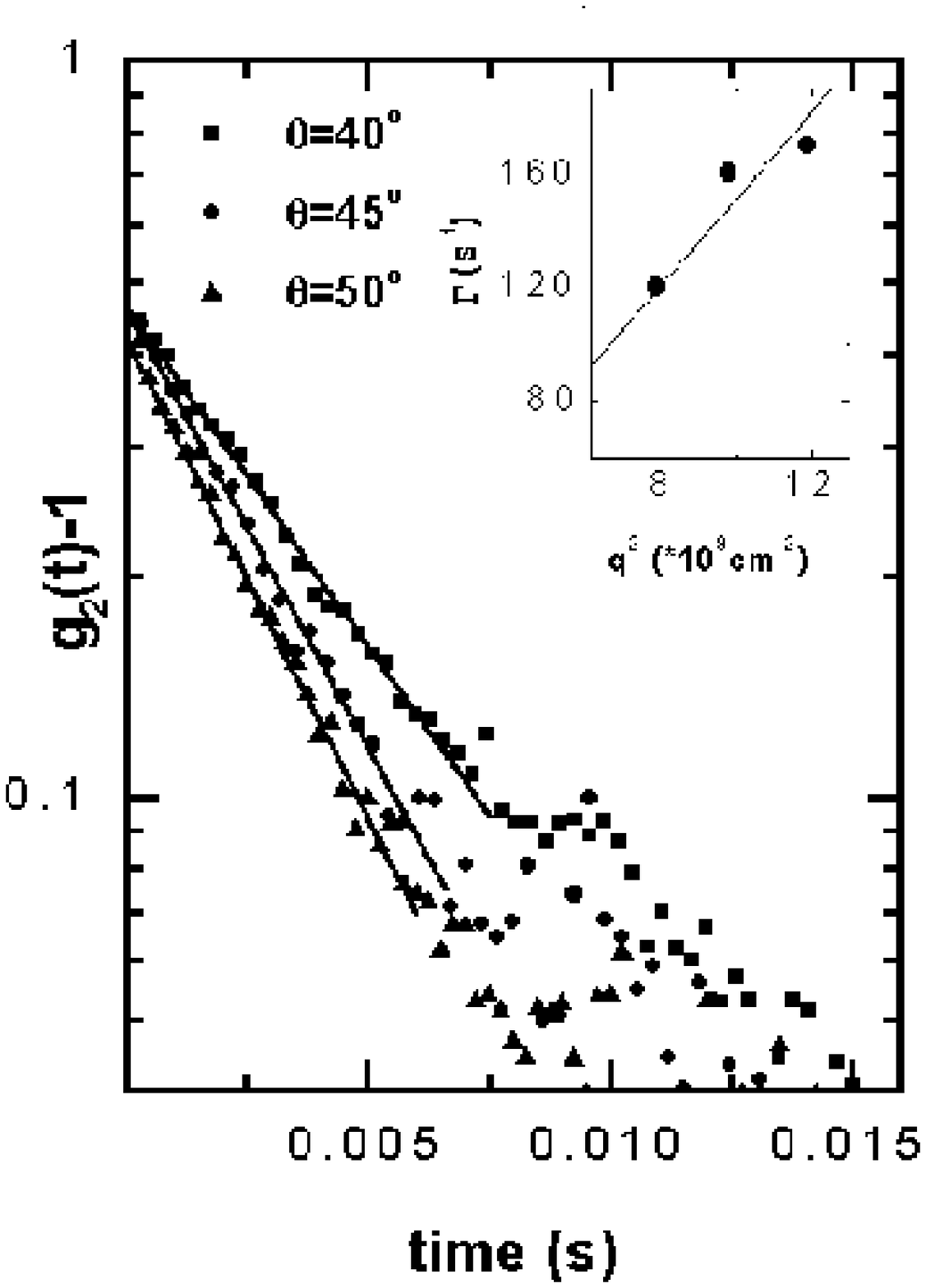}}
\label{Fig. 1}
\end{figure}
\flushbottom{\bf{Fig. 1}}

\begin{figure}
\centerline{\epsfxsize = 16cm \epsfbox{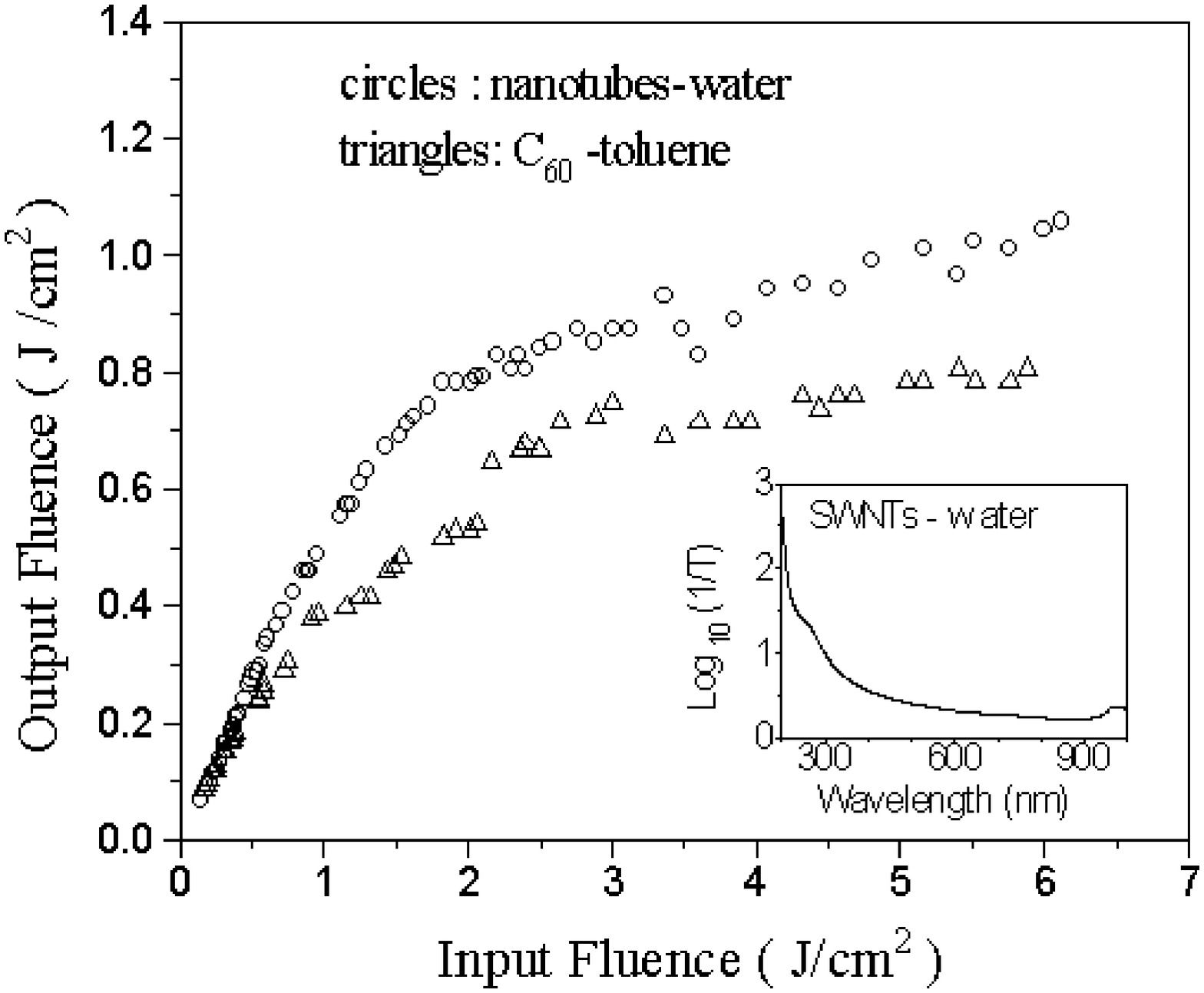}}
\label{Fig. 2}
\end{figure}
\flushbottom{\bf{Fig. 2}}

\begin{figure}
\centerline{\epsfxsize = 16cm \epsfbox{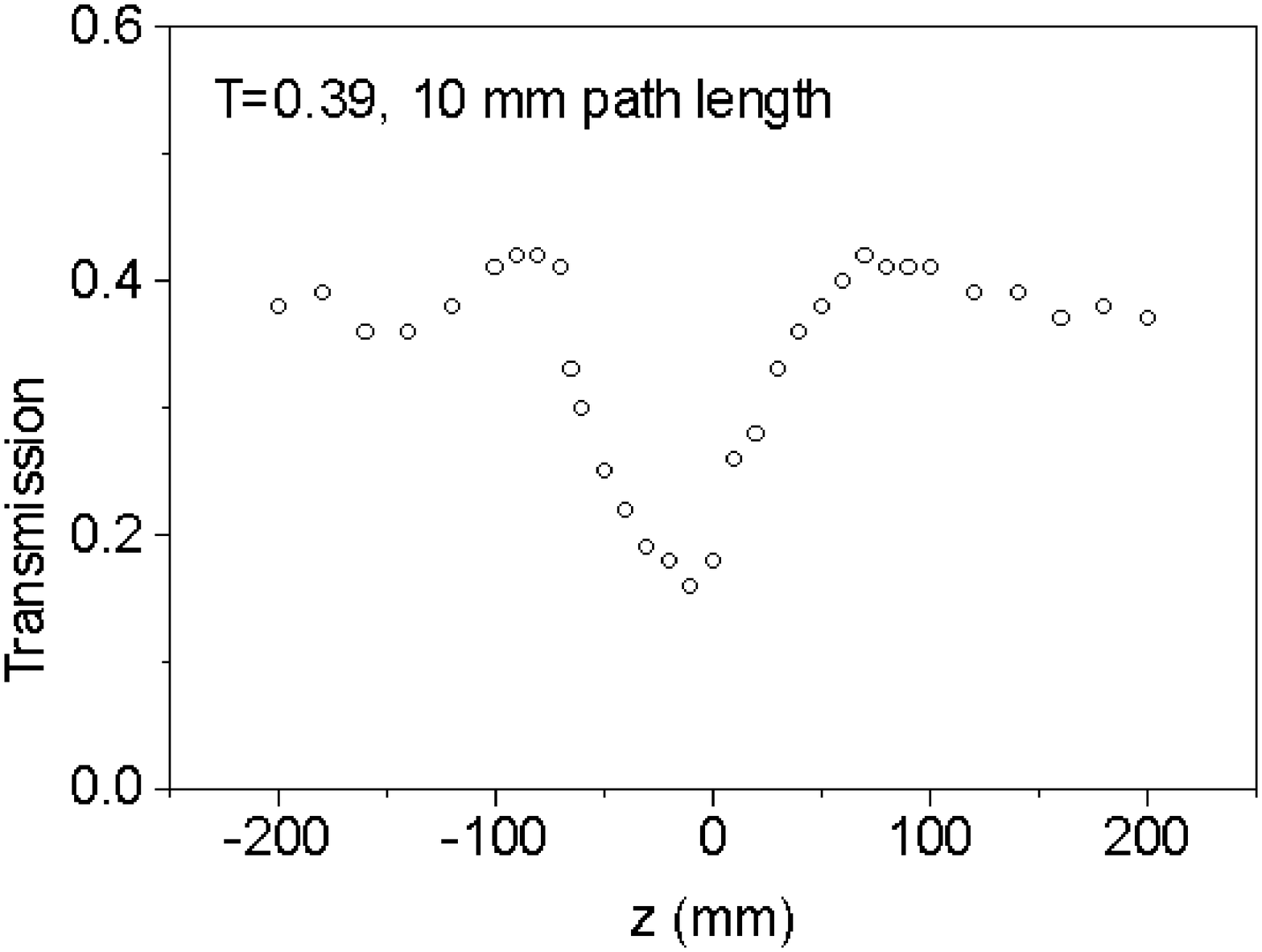}}
\label{Fig. 3}
\end{figure}
\flushbottom{\bf{Fig. 3}}

\begin{figure}
\centerline{\epsfxsize = 16cm \epsfbox{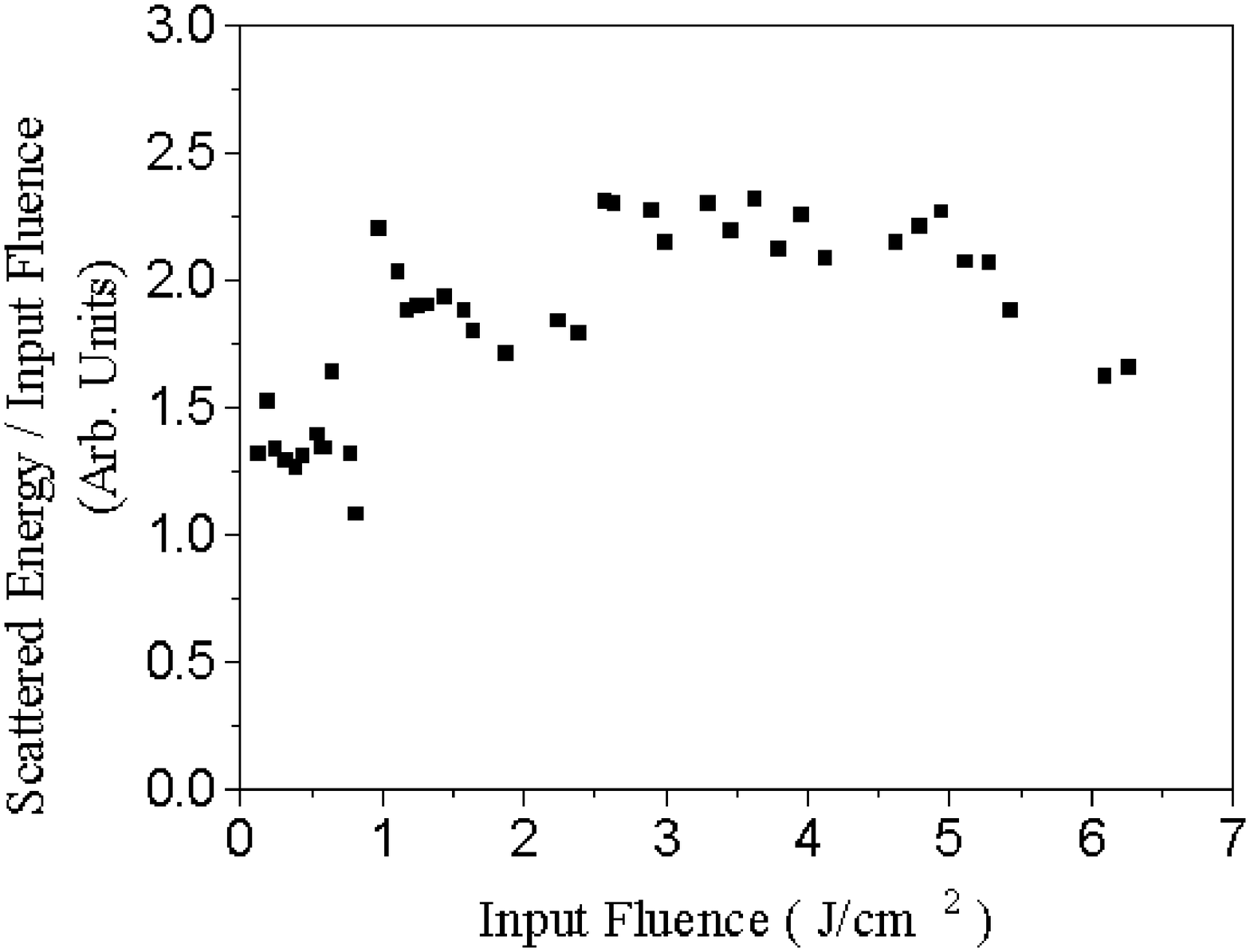}}
\label{Fig. 4}
\end{figure}
\flushbottom{\bf{Fig. 4}}

\begin{figure}
\centerline{\epsfxsize = 20cm \epsfbox{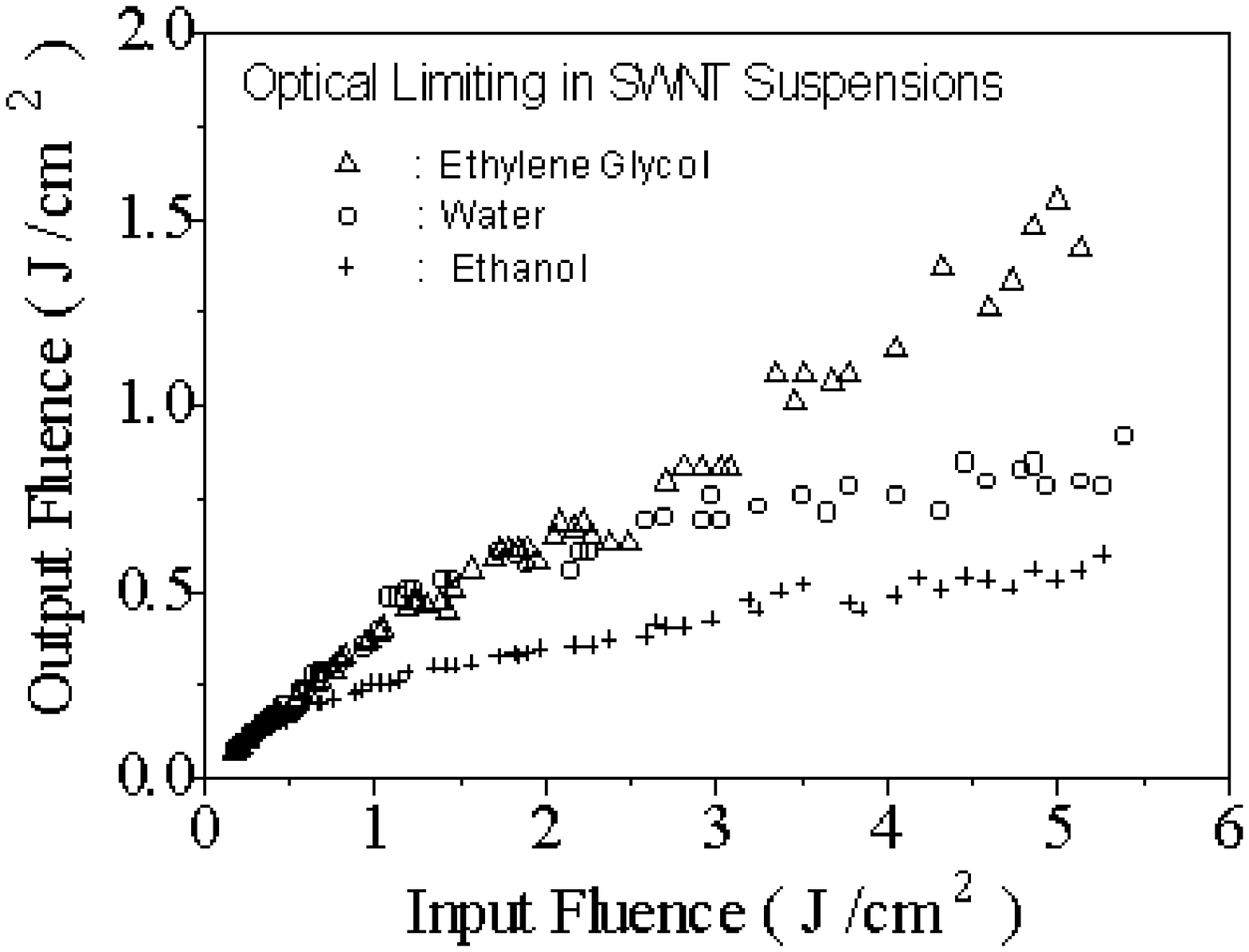}}
\label{Fig. 5}
\end{figure}
\flushbottom{\bf{Fig. 5}} \end{document}